\documentclass[aps,prc,superscriptaddress,reprint]{revtex4-2}
\usepackage{graphicx}
\usepackage{multirow}
\usepackage[dvipsnames]{xcolor}
\usepackage{amssymb}
\usepackage{amsmath}
\usepackage{bm}
\usepackage{lineno}
\usepackage{booktabs}
\usepackage[colorlinks=true, breaklinks=true, linkcolor=blue, citecolor=violet, urlcolor=cyan]{hyperref}
\usepackage{placeins}

\renewcommand{\Im}{\mathop{\mathrm{Im}}}

\newcommand{\diff}{{\mathrm{d}}}
\newcommand{\nuc}[2]{\relax\ifmmode{}^{#1}{\protect\text{#2}}\else${}^{#1}$#2\fi}

\begin{document}

\title{Exterior complex scaling enables physics-informed neural networks for nuclear reactions}

\author{Jin Lei}
\email[]{jinl@tongji.edu.cn}
\affiliation{School of Physics Science and Engineering, Tongji University, Shanghai 200092, China.}
\affiliation{Southern Center for Nuclear-Science Theory (SCNT), Institute of Modern Physics, Chinese Academy of Sciences, Huizhou 516000, Guangdong Province, China.}

\begin{abstract}
Physics-informed neural networks (PINNs) have emerged as a powerful tool for solving differential equations, yet their application to nuclear scattering has been hindered by the oscillatory, non-decaying nature of scattering wave functions. In this work, I demonstrate that exterior complex scaling (ECS) transforms scattering boundary conditions into exponentially decaying waves suitable for neural network solutions, enabling PINNs to solve nuclear reaction problems for the first time. I develop a driven-equation formulation where the source term is confined to the real axis, avoiding the need to analytically continue nuclear potentials into the complex plane. The method is validated on nucleon-nucleus scattering (n+$^{40}$Ca at $E_{\text{lab}}=20$~MeV) with 21 partial waves, achieving phase shift accuracy of $\Delta\delta \lesssim 0.1^\circ$ for the strongly absorbed channels ($\ell \leq 4$) and $\Delta\delta \leq 0.60^\circ$ for all channels up to $\ell = 10$, when compared to conventional solvers. I further demonstrate the approach on heavy-ion scattering ($^6$Li+$^{208}$Pb at 40~MeV) with 41 partial waves and strong Coulomb effects, where an auto-adaptive anchor warm-down for weak-source channels yields a mean S-matrix accuracy of $|\Delta|S_\ell|| \approx 3 \times 10^{-3}$ across the full angular momentum range, including the absorption-to-transparency transition region. This work establishes the foundation for extending PINNs to inverse problems where end-to-end differentiability enables direct fitting of optical potential parameters, coupled-channel reactions, and few-body scattering where traditional grid methods face exponential scaling.
\end{abstract}


\date{\today}

\maketitle

\section{Introduction}
\label{sec:intro}

Nuclear scattering experiments provide fundamental information about nuclear structure and reaction mechanisms that underpin current understanding of atomic nuclei. When a beam of particles strikes a target nucleus, the angular distribution of scattered particles encodes information about the nuclear potential through which the projectile propagates. This potential, known as the optical model potential, contains both real components describing elastic scattering and imaginary components accounting for absorption into non-elastic channels such as compound nucleus formation, direct reactions, and particle emission. Extracting nuclear structure information from scattering data requires solving the quantum mechanical scattering problem, typically formulated as a partial-wave decomposition of the Schr\"{o}dinger equation with appropriate boundary conditions at infinity~\cite{Frobrich1996,Thompson2009}.

The computational challenge of nuclear scattering lies not in the differential equation itself, which is a straightforward second-order ordinary differential equation, but rather in the boundary conditions. Unlike bound-state problems where the wave function decays exponentially at large distances, scattering wave functions remain oscillatory at all radii, behaving asymptotically as linear combinations of incoming and outgoing spherical waves. Traditional numerical methods must propagate solutions to sufficiently large radii where the nuclear potential vanishes, then match to known asymptotic forms to extract phase shifts and S-matrix elements. For systems with long-range Coulomb interactions, this matching becomes particularly delicate because the Coulomb potential extends to infinity and modifies the asymptotic behavior in nontrivial ways involving logarithmic phase corrections.

Physics-informed neural networks have recently emerged as a powerful paradigm for solving differential equations by training neural networks to minimize residuals of the governing equations at collocation points~\cite{Lagaris1998,Raissi2019,Karniadakis2021}. Unlike traditional numerical methods that discretize the computational domain, PINNs represent solutions as continuous, differentiable functions parameterized by network weights. This approach has proven successful for a wide range of problems in fluid mechanics, heat transfer, and solid mechanics, particularly for stationary problems. More broadly, neural networks have demonstrated remarkable capability in solving quantum mechanical problems, including variational solutions of the many-body Schr\"{o}dinger equation~\cite{Carleo2017,Hermann2020,Pfau2020}. The key advantage of PINNs is their mesh-free nature and ability to incorporate physics constraints directly into the loss function, making them attractive for problems with complex geometries or those where generating high-quality meshes is challenging. Libraries such as DeepXDE~\cite{Lu2021} have made these methods increasingly accessible.

Despite their success in many domains, PINNs face a fundamental obstacle when applied to scattering problems. Scattering wave functions satisfy radiation boundary conditions (Sommerfeld conditions) at infinity, requiring the solution to be a specific superposition of incoming and outgoing waves. These conditions cannot be reduced to simple Dirichlet or Neumann conditions at any finite boundary because the wave function oscillates indefinitely and the correct asymptotic form involves the unknown phase shift that one is trying to compute. Truncating the computational domain at any finite radius introduces spurious reflections that contaminate the solution. This fundamental incompatibility between oscillatory radiation conditions and finite-domain computation has prevented the application of PINNs to nuclear reactions.

Exterior complex scaling provides an elegant solution to this boundary condition problem by analytically continuing spatial coordinates into the complex plane beyond a certain radius~\cite{Simon1979,Reinhardt1982,Ho1983,Moiseyev1998}. When the radial coordinate $r$ is rotated into the complex plane by an angle $\theta$, outgoing waves that behave as $e^{ikr}$ become exponentially damped as $e^{ikr\cos\theta}e^{-kr\sin\theta}$. This transformation converts the scattering problem into one with square-integrable ($L^2$) boundary conditions: the wave function now decays exponentially at large distances and can be set to zero at a finite outer boundary, serving a similar purpose to perfectly matched layer absorbing boundaries~\cite{Berenger1994} that have also been adapted for quantum wave equations~\cite{Antoine2017}. The physics of the scattering process is preserved because the complex scaling only affects the asymptotic region where the potential vanishes, leaving the nuclear interaction region on the real axis where wave functions and observables are physically meaningful.

The application of complex scaling to scattering calculations was pioneered in atomic and molecular physics for studying resonances and electron-atom collisions~\cite{Rescigno2000}. In nuclear physics, complex scaling has been applied to study resonances in light nuclei~\cite{Myo2020}, and Liu, Lei, and Ren recently developed the COLOSS code implementing complex scaling for optical model calculations~\cite{Liu2025}. Conventional discretization methods for scattering, such as finite differences, R-matrix~\cite{Wigner1947,Lane1958,Descouvemont2010}, and Numerov integration~\cite{Thorlacius1987}, are highly efficient for single-channel scattering, and the present work does not aim to replace them for such applications. Rather, the motivation for combining ECS with PINNs lies in the unique advantages that neural network approaches offer for more challenging problems.

The key advantages of a PINN-based scattering solver are threefold. First, the entire computational pipeline is end-to-end differentiable, enabling gradient-based optimization of potential parameters to fit experimental data, i.e., inverse problems that would otherwise require expensive outer-loop optimization over repeated forward solves. Second, PINNs naturally extend to multi-output representations, making coupled-channel calculations conceptually straightforward without explicitly constructing coupled matrix equations. Third, mesh-free neural network methods scale more favorably with dimensionality than grid-based approaches, offering a potential pathway to few-body scattering problems (e.g., Faddeev equations~\cite{Faddeev1961,Faddeev1993,Lazauskas2019}) where traditional methods face exponential computational cost. The present work establishes the foundation by validating PINN-ECS against established solvers on well-understood single-channel benchmarks.

I develop a driven-equation formulation that decomposes the wave function into a known incident Coulomb wave and a scattered wave satisfying an inhomogeneous equation whose source term is confined to the real axis (see Sec.~\ref{sec:theory} for details). Several technical innovations, including a sigmoid-capped boundary condition factor for high-$\ell$ channels, an auto-adaptive anchor scale with a linear warm-down for weak-source channels (high-$\ell$ partial waves where the centrifugal barrier suppresses the source term), and multi-point S-matrix averaging, are described in Sec.~\ref{sec:method}.

In this paper, I validate the PINN-ECS method on two benchmark systems. The first is neutron scattering on $^{40}$Ca at $E_{\text{lab}} = 20$~MeV using the Koning-Delaroche global optical potential, which includes volume, surface, and spin-orbit terms~\cite{Koning2003}. I compute 21 partial waves up to $\ell = 10$ and compare phase shifts and S-matrix elements to those obtained from the established COLOSS code. The second benchmark is $^6$Li+$^{208}$Pb elastic scattering at 40~MeV, a heavy-ion system with strong Coulomb effects requiring 41 partial waves. I compute the Rutherford ratio $(\diff\sigma/\diff\Omega)/(\diff\sigma/\diff\Omega)_{\text{Ruth}}$ and demonstrate the characteristic Coulomb-nuclear interference pattern.

The paper is organized as follows. Section~\ref{sec:theory} presents the theoretical framework, including the radial Schr\"{o}dinger equation, the exterior complex scaling transformation, and the driven-equation formulation for scattering. Section~\ref{sec:method} describes the neural network architecture, loss functions, and training procedure. Section~\ref{sec:results} presents benchmark results for n+$^{40}$Ca and $^6$Li+$^{208}$Pb scattering. Section~\ref{sec:discussion} discusses the capabilities and limitations of the approach. Section~\ref{sec:conclusions} summarizes my findings and outlines future directions.

\section{Theoretical Framework}
\label{sec:theory}

The quantum mechanical description of nuclear scattering begins with the time-independent Schr\"{o}dinger equation for the relative motion of two colliding nuclei. In the center-of-mass frame, the three-dimensional wave function can be expanded in partial waves, each characterized by orbital angular momentum quantum number $\ell$. For each partial wave, the radial wave function $u_\ell(r) = r R_\ell(r)$, where $R_\ell(r)$ is the radial part of the full wave function, satisfies a one-dimensional Schr\"{o}dinger equation that forms the starting point of my analysis.

The radial Schr\"{o}dinger equation in the variable $x$ (the physical radial coordinate) takes the form
\begin{equation}
\left[-\frac{\hbar^2}{2\mu}\frac{\diff^2}{\diff x^2} + \frac{\hbar^2\ell(\ell+1)}{2\mu x^2} + V_N(x) + V_C(x)\right] u(x) = E\, u(x),
\label{eq:schrodinger}
\end{equation}
where $\mu$ is the reduced mass of the projectile-target system, $E$ is the center-of-mass kinetic energy, $V_N(x)$ is the short-range nuclear potential, and $V_C(x)$ is the Coulomb potential arising from the electrostatic repulsion between nuclei with charges $Z_1$ and $Z_2$, modeled as a uniformly charged sphere with radius $R_C = r_C A^{1/3}$ (using $r_C = 1.3$~fm) to avoid the point-charge singularity at the origin. The centrifugal term $\hbar^2\ell(\ell+1)/(2\mu x^2)$ arises from the angular part of the kinetic energy operator and creates an effective barrier that prevents low-energy particles from penetrating to small radii for high angular momenta. The wave function must satisfy the boundary condition $u(0) = 0$ to remain finite at the origin, and for scattering states at positive energy $E > 0$, it must match the appropriate asymptotic form at large distances.

The nuclear potential $V_N(x)$ is the optical model potential~\cite{Feshbach1958,Feshbach1962,Satchler1983}, a short-range complex function whose imaginary part accounts for flux loss to non-elastic channels. The detailed parametrizations used in the present benchmarks (Koning-Delaroche~\cite{Koning2003} for n+$^{40}$Ca and a Cook-type Woods-Saxon volume potential~\cite{Cook1982} for $^6$Li+$^{208}$Pb) are specified in Sec.~\ref{sec:results} where the corresponding calculations are presented.

For scattering states, the asymptotic behavior of the wave function at large $x$ where $V_N(x) \to 0$ is determined by the Coulomb potential alone. A convenient standard normalization writes the asymptotic solution as
\begin{equation}
u(x) \xrightarrow{x \to \infty} \frac{i}{2}\left[H_\ell^{(-)}(\eta, kx) - S_\ell\, H_\ell^{(+)}(\eta, kx)\right],
\label{eq:asymptotic}
\end{equation}
where $k = \sqrt{2\mu E}/\hbar$ is the wave number, $\eta = Z_1 Z_2 e^2 \mu/(\hbar^2 k)$ is the Sommerfeld parameter characterizing the strength of the Coulomb interaction, and $H_\ell^{(\pm)} = G_\ell \pm i F_\ell$ are the outgoing (+) and incoming ($-$) Coulomb-Hankel functions constructed from the regular ($F_\ell$) and irregular ($G_\ell$) Coulomb wave functions~\cite{Abramowitz1964,Thompson1985}. The S-matrix element $S_\ell$ therefore multiplies the outgoing Coulomb-Hankel function relative to the incoming one in this asymptotic normalization. For purely elastic scattering with a real potential, $|S_\ell| = 1$ and $S_\ell = e^{2i\delta_\ell}$; with absorption one may write $S_\ell = |S_\ell| e^{2i\delta_\ell}$, where $|S_\ell| < 1$ measures the flux loss to non-elastic channels~\cite{Taylor1972,Newton1982}.

The traditional approach to solving the scattering problem is to integrate Eq.~(\ref{eq:schrodinger}) outward from the origin using methods such as Numerov~\cite{Pillai2012} or Runge-Kutta, imposing the boundary condition $u(0) = 0$ and the correct $r^{\ell+1}$ behavior near the origin. The integration continues to a matching radius $r_m$ sufficiently large that the nuclear potential has decayed to negligible values, typically $r_m \approx 15$--20~fm for nucleon-nucleus scattering. At this point, the numerical solution is matched to the known asymptotic form in Eq.~(\ref{eq:asymptotic}) to extract $S_\ell$. For neutral particles ($\eta = 0$), the Coulomb functions reduce to spherical Bessel functions, while for charged particles, specialized algorithms such as COULCC~\cite{Thompson1985} are required to evaluate them accurately.

This conventional approach, while reliable, does not extend naturally to neural network solvers because of the radiation boundary conditions at large $r$. Physics-informed neural networks operate on finite computational domains and require boundary conditions that can be specified independently of the solution. Scattering problems, however, require Sommerfeld radiation conditions: the wave function must asymptotically match a superposition of incoming and outgoing waves whose relative amplitude (the S-matrix) is the unknown quantity being computed. These oscillatory conditions cannot be imposed at any finite boundary, and truncating the domain introduces spurious reflections. This fundamental incompatibility has prevented the application of PINNs to scattering problems.

Exterior complex scaling resolves this difficulty by analytically continuing the radial coordinate into the complex plane beyond a scaling radius $R_0$~\cite{Simon1979}. I introduce a smooth coordinate mapping from a parameter coordinate $r \in [0, R_{\text{max}}]$ (always real) to a physical coordinate $x(r)$ that becomes complex for $r > R_0$. The mapping is constructed using a smooth switching function
\begin{equation}
s(t) = \begin{cases}
0 & t \leq 0 \\
3t^2 - 2t^3 & 0 < t < 1 \\
1 & t \geq 1
\end{cases},
\quad t = \frac{r - R_0}{w},
\label{eq:switching}
\end{equation}
where $w$ is the width of the transition region, typically chosen as $w \approx 0.1 R_0$ to ensure a smooth but localized transition. This polynomial form ensures that $s(t)$ and its first derivative are continuous everywhere, avoiding discontinuities that could cause numerical difficulties. The physical coordinate is then
\begin{equation}
x(r) = r + (e^{i\theta} - 1)\, I(r),
\label{eq:ecs_coord}
\end{equation}
where $\theta$ is the rotation angle and $I(r)$ is the integral of the switching function,
\begin{equation}
I(r) = \int_0^r s\left(\frac{r'-R_0}{w}\right) \diff r'.
\end{equation}
This integral has a closed-form expression: $I(r) = 0$ for $r \leq R_0$, $I(r) = w(t^3 - t^4/2)$ for $R_0 < r < R_0 + w$, and $I(r) = w/2 + (r - R_0 - w)$ for $r \geq R_0 + w$. The Jacobian of the transformation is
\begin{equation}
q(r) = \frac{\diff x}{\diff r} = 1 + (e^{i\theta} - 1)\, s\left(\frac{r-R_0}{w}\right),
\label{eq:jacobian}
\end{equation}
which equals unity for $r \leq R_0$ (where $x = r$ is real) and approaches $e^{i\theta}$ for $r \geq R_0 + w$ (where $x$ increases along a ray at angle $\theta$ in the complex plane).

The effect of this coordinate transformation on outgoing waves is crucial. An outgoing wave $e^{ikx}$ in the ECS region becomes $e^{ikx(r)} = e^{ik[r\cos\theta + iy(r)]}$, where $y(r) = \Im[x(r)]$ increases approximately as $(r-R_0)\sin\theta$ for $r > R_0 + w$. The amplitude therefore decays as $e^{-k(r-R_0)\sin\theta}$, converting the oscillatory wave into an exponentially damped function. For the ECS absorbing boundary to work effectively, the damping factor $\exp[-k(R_{\text{max}}-R_0)\sin\theta]$ at the outer boundary must be small enough to suppress the wave function there; in practice I require it to be below $\approx 10^{-3}$. The specific parameter values $\theta$, $R_0$, $R_{\text{max}}$ used for each benchmark system, together with the corresponding numerical decay factors, are quoted in Sec.~\ref{sec:results} where the calculations are presented.

Unlike uniform complex scaling where the nuclear potential must be evaluated at complex coordinates, ECS keeps the nuclear interaction region on the real axis by choosing $R_0$ beyond the range of $V_N$~\cite{McCurdy2004}. Since the nuclear potential vanishes for $r > R_0$, the Woods-Saxon singularity structure does not constrain the rotation angle $\theta$. Only the Coulomb potential $V_C \propto 1/x(r)$, which is analytic away from the origin, requires evaluation at complex coordinates. A moderate $\theta$ provides adequate absorption while avoiding numerical difficulties in evaluating Coulomb wave functions at complex arguments; larger values of $\theta$ would require more collocation points in the ECS region to resolve the more rapidly decaying solution.

Under the ECS coordinate transformation, the differential operators transform according to the chain rule. The first derivative becomes $\diff/\diff x = q^{-1}\,\diff/\diff r$, and the second derivative is
\begin{equation}
\frac{\diff^2}{\diff x^2} = \frac{1}{q^2}\frac{\diff^2}{\diff r^2} - \frac{q'}{q^3}\frac{\diff}{\diff r},
\label{eq:ecs_operators}
\end{equation}
where $q' = \diff q/\diff r = (e^{i\theta}-1)(6t-6t^2)/w$ in the transition region $0 < t < 1$, and $q' = 0$ outside this region. The transformed Schr\"{o}dinger equation in the parameter coordinate $r$ takes the form
\begin{equation}
-\frac{\hbar^2}{2\mu}\left(\frac{1}{q^2}u'' - \frac{q'}{q^3}u'\right) + \frac{\hbar^2\ell(\ell+1)}{2\mu\,x(r)^2}u + V(x(r))\,u = E\,u,
\label{eq:ecs_schrodinger}
\end{equation}
where primes denote derivatives with respect to $r$ and $V(x) = V_N(x) + V_C(x)$ is the total potential. The critical simplification is that for $r > R_0$, I set $V_N = 0$ because the nuclear potential has negligible range beyond the nuclear surface. Only the Coulomb potential requires evaluation at complex coordinates, and this is straightforward since $1/x$ is analytic everywhere except at the origin.

For the scattering problem, I employ a driven-equation formulation that separates the total wave function into incident and scattered components,
\begin{equation}
u(r) = u^{\text{in}}(r) + u^{\text{sc}}(r),
\label{eq:decomposition}
\end{equation}
where $u^{\text{in}}(r) = F_\ell(\eta, kx(r))$ is the regular point-Coulomb wave function used as the reference solution, computed with the point-Coulomb Sommerfeld parameter $\eta$. The scattered wave $u^{\text{sc}}(r)$ satisfies the inhomogeneous driven equation
\begin{equation}
(H - E)\, u^{\text{sc}}(r) = -V_{\text{short}}(r)\, u^{\text{in}}(r),
\label{eq:driven}
\end{equation}
where $H$ is the Hamiltonian operator including kinetic energy, centrifugal barrier, the charged-sphere Coulomb potential, and the nuclear optical potential. The short-range potential $V_{\text{short}} = V_N + (V_C^{\text{sphere}} - V_C^{\text{point}})$ includes both the nuclear potential and the charged-sphere correction; the latter arises because the incident Coulomb wave is a solution of the point-Coulomb Schr\"{o}dinger equation, not the charged-sphere equation. The charged-sphere correction vanishes identically for $r > R_C$, while the nuclear term becomes negligible well before the chosen ECS radius $R_0$ and is set to zero in the ECS region. This means the right-hand side of Eq.~(\ref{eq:driven}) is confined entirely to the real-axis region $r \leq R_0$, and no analytic continuation of $V_{\text{short}}$ is required. In the ECS region, the scattered wave satisfies the homogeneous equation $(H-E)u^{\text{sc}} = 0$ with exponentially decaying boundary conditions, which is precisely the type of problem that neural networks can solve efficiently.

The S-matrix element is extracted by matching the total wave function at a radius $r_m < R_0$ on the real axis. The matching uses Wronskians of $u$ with the regular and irregular Coulomb functions $F_\ell(\eta,\rho)$, $G_\ell(\eta,\rho)$:
\begin{align}
N &= k\, u\, F_\ell' - u'\, F_\ell, \\
D &= u'\, G_\ell - k\, u\, G_\ell',
\end{align}
where all functions are evaluated at $r = r_m$ ($\rho_m = k r_m$), $u' = \diff u/\diff r$, and $F'_\ell$, $G'_\ell$ denote derivatives with respect to $\rho = kr$. The S-matrix is then computed directly as
\begin{equation}
S_\ell = \frac{D + iN}{D - iN},
\label{eq:smatrix}
\end{equation}
which remains well-defined for absorptive optical potentials and avoids the numerical pathology of the $K$-matrix representation when $D \to 0$ (corresponding to $\delta_\ell \to \pm 90^\circ$). The phase shift is then $\delta_\ell = \arg(S_\ell)/2$ and the absorption is characterized by $|S_\ell| \leq 1$. To improve robustness, I average the S-matrix over five matching radii distributed uniformly in the interval $[0.7 R_0, 0.9 R_0]$, which reduces sensitivity to local fluctuations in the numerical solution.

\section{Computational Method}
\label{sec:method}

The PINN-ECS method combines exterior complex scaling, which transforms oscillatory scattering waves into decaying functions, with a physics-informed neural network that learns the scattered wave function by minimizing the differential equation residual. The key components include the ECS coordinate mapping, the network architecture with hard-coded boundary conditions, and the driven-equation formulation that confines the nuclear potential to the real axis.

The physics-informed neural network represents the scattered wave function $u^{\text{sc}}(r)$ as a parameterized function of the radial coordinate. The network architecture consists of a fully-connected feedforward network with $r \in [0, R_{\text{max}}]$ as input and two outputs representing the real and imaginary parts of the complex wave function, $u_R(r)$ and $u_I(r)$. I use four hidden layers with 128 neurons each and sinusoidal activation functions, $\sigma(z) = \sin(z)$, which are particularly well-suited for representing oscillatory solutions~\cite{Sitzmann2020,Jagtap2020}. The raw network output $\hat{u}(r)$ is then modified to enforce boundary conditions exactly.

\begin{figure*}[t]
\centering
\includegraphics[width=\textwidth]{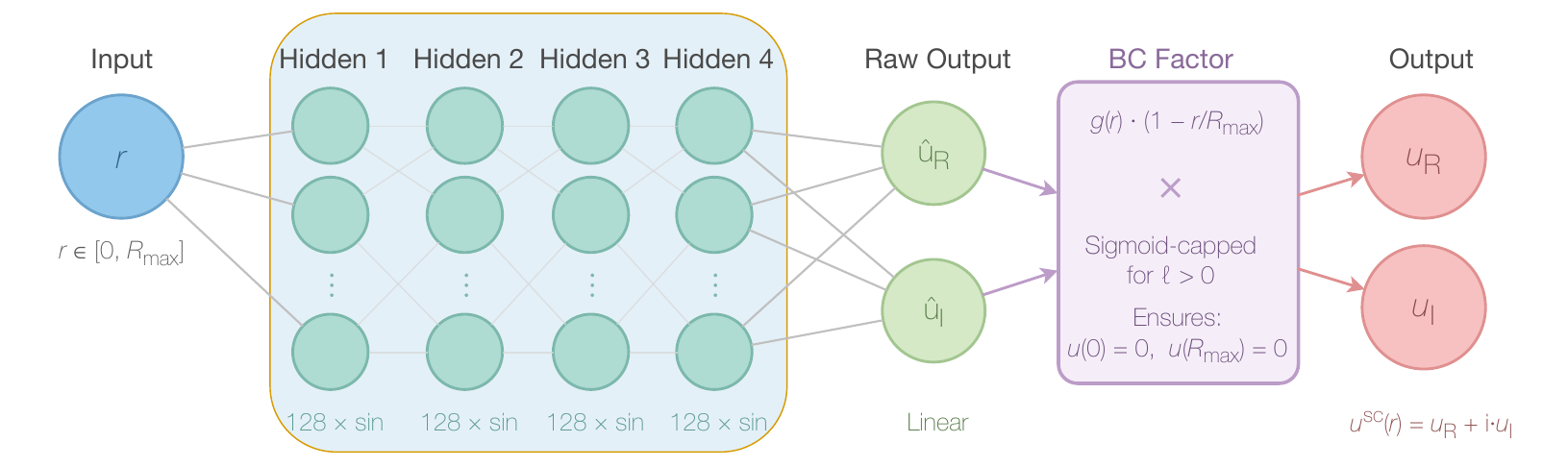}
\caption{Neural network architecture for PINN-ECS. The network takes the radial coordinate $r$ as input and passes it through four hidden layers of 128 neurons each with sinusoidal activation functions. The raw network output $\hat{u}(r)$ is multiplied by the boundary condition factor $g(r)\,(1 - r/R_{\text{max}})$, where $g(r) = (1-\sigma_c)\,r^{\ell+1} + \sigma_c\,R_{\text{nuc}}^{\ell+1}$ [Eq.~(\ref{eq:sigmoid_cap})] is a sigmoid-capped factor that prevents numerical overflow for high angular momentum channels. This hard-coded construction ensures $u^{\text{sc}}(0) = u^{\text{sc}}(R_{\text{max}}) = 0$ exactly, regardless of the learned weights.}
\label{fig:nn_architecture}
\end{figure*}

The boundary conditions $u^{\text{sc}}(0) = 0$ and $u^{\text{sc}}(R_{\text{max}}) = 0$ are enforced through a multiplicative factor,
\begin{equation}
u^{\text{sc}}(r) = \frac{g(r)\,(1 - r/R_{\text{max}})}{\mathcal{N}}\, \hat{u}(r),
\label{eq:bc_factor}
\end{equation}
where $\mathcal{N}$ is a normalization constant and $g(r)$ encodes the correct $r^{\ell+1}$ behavior near the origin. A naive choice $g(r) = r^{\ell+1}$ becomes problematic at high angular momentum. Since the final wave function is the product $u^{\text{sc}} \propto g(r) \cdot \hat{u}(r)$ and must remain bounded, a large $g(r)$ requires the network to output proportionally small values of $\hat{u}(r)$. For example, at $\ell = 5$ and $r = R_0 = 20$~fm, the factor $r^{\ell+1} \approx 6 \times 10^7$. To produce a bounded wave function, the optimizer must drive all network weights toward small values, which compresses the dynamic range available for representing the detailed structure in the nuclear interior.

\begin{figure}[t]
\centering
\includegraphics[width=\columnwidth]{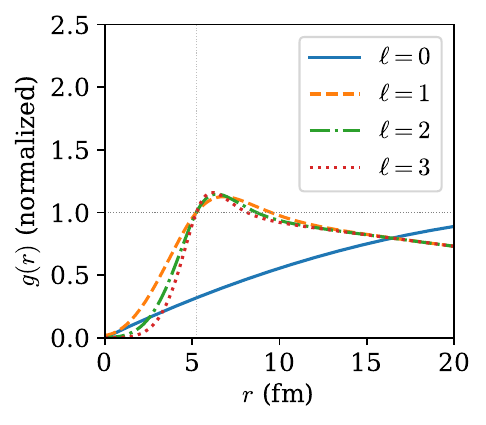}
\caption{Normalized boundary condition factor $g(r)$ showing the sigmoid capping that prevents growth beyond $R_{\text{nuc}}^{\ell+1}$: $\ell = 0$ (solid), $\ell = 1$ (dashed), $\ell = 2$ (dash-dotted), and $\ell = 3$ (dotted). The vertical dashed line marks $R_{\text{nuc}}$. For $\ell > 0$, the factor transitions smoothly from $r^{\ell+1}$ behavior near the origin to a bounded value $R_{\text{nuc}}^{\ell+1}$ in the asymptotic region.}
\label{fig:bc_engineering}
\end{figure}

I resolve this problem with a sigmoid-capped boundary condition factor,
\begin{equation}
g(r) = (1-\sigma_c)\, r^{\ell+1} + \sigma_c\, R_{\text{nuc}}^{\ell+1},
\label{eq:sigmoid_cap}
\end{equation}
where $R_{\text{nuc}}$ is a characteristic nuclear radius (I use $R_{\text{nuc}} = 1.25 A_{\text{targ}}^{1/3}$~fm, consistent with the radius parametrization $R = r\, A_{\text{targ}}^{1/3}$ of the KD02 and Cook optical potentials used in Sec.~\ref{sec:results}) and $\sigma_c$ is a smooth transition function,
\begin{equation}
\sigma_c(r) = \left[1 + \exp\left(-\frac{2(\ell+1)}{R_{\text{nuc}}}(r - R_{\text{nuc}})\right)\right]^{-1}.
\end{equation}
This construction ensures that $g(r) \approx r^{\ell+1}$ for $r \ll R_{\text{nuc}}$ (preserving the correct behavior at the origin) while $g(r) \to R_{\text{nuc}}^{\ell+1}$ for $r \gg R_{\text{nuc}}$ (bounded in the asymptotic region). Figure~\ref{fig:nn_architecture} shows how this BC factor is integrated into the neural network architecture, where the raw network output is multiplied by $g(r)$ to enforce the boundary conditions exactly. Figure~\ref{fig:bc_engineering} illustrates how this capping prevents the pathological growth for high-$\ell$ channels. The normalized BC factor $g(r)$ is shown for $\ell = 0$ (solid), $\ell = 1$ (dashed), $\ell = 2$ (dash-dotted), and $\ell = 3$ (dotted). The vertical dashed line marks $R_{\text{nuc}}$. For $\ell > 0$, the factor rises steeply near the origin following $r^{\ell+1}$, then saturates to a constant beyond $R_{\text{nuc}}$. For $\ell = 0$, where the uncapped factor $r(1-r/R_{\text{max}})$ is already well-behaved, I skip the sigmoid capping.

The loss function combines the differential equation residual with an anchor term that prevents the trivial solution $u^{\text{sc}} = 0$. The residual loss is
\begin{equation}
\mathcal{L}_{\text{res}} = \frac{1}{N_c}\sum_{i=1}^{N_c} \left|[(H-E)u^{\text{sc}} + V_{\text{short}} u^{\text{in}}](r_i)\right|^2,
\label{eq:loss_res}
\end{equation}
where the sum runs over $N_c$ collocation points $\{r_i\}$ and the Hamiltonian action is computed using automatic differentiation~\cite{Baydin2018} for the scattered wave derivatives. The anchor loss prevents the optimizer from finding the trivial solution $u^{\text{sc}} = 0$,
\begin{equation}
\mathcal{L}_{\text{anchor}} = w_a \exp\left(-\frac{\langle|u^{\text{sc}}|^2\rangle}{s_a}\right),
\label{eq:loss_anchor}
\end{equation}
where $\langle \cdot \rangle$ denotes the mean over collocation points, $w_a$ is the anchor weight (set to $w_a = 0.1$ throughout this work), and $s_a$ is a scale parameter. When $u^{\text{sc}} \approx 0$, the exponential approaches unity and the loss receives a penalty of $w_a$; when $|u^{\text{sc}}|^2 \gg s_a$, the anchor loss vanishes and does not interfere with residual minimization. The total loss is $\mathcal{L} = \mathcal{L}_{\text{res}} + \mathcal{L}_{\text{anchor}}$. The scale parameter is set automatically based on the source term magnitude:
\begin{equation}
s_a = \langle |V_{\text{short}} F_\ell|^2 \rangle,
\label{eq:auto_anchor}
\end{equation}
which ensures that the anchor contribution remains comparable to the residual loss across all partial waves and eliminates the need for manual tuning. For heavy-ion scattering, partial waves in the absorption-to-transparency transition region ($\ell \approx 25$--$30$ for the $^6$Li+$^{208}$Pb benchmark) pose a special challenge: the source term is very weak (centrifugal suppression) and the correct scattered wave is itself very small because $|S_\ell|$ approaches unity, so any spurious anchor contribution inflates $|u^{\text{sc}}|$ above its physical value and biases $|S_\ell|$ systematically below unity (i.e.\ produces apparent absorption that is not really there). To remove this bias I introduce a linear anchor warm-down for channels with $s_a < s_a^* = 0.1$:
\begin{equation}
\mathcal{L}_{\text{anchor}}(n) \to \max\!\left(0,\, 1 - \frac{2n}{N_{\text{wd}}}\right)\,\mathcal{L}_{\text{anchor}},
\label{eq:warmdown}
\end{equation}
where $n$ is the current Adam epoch and $N_{\text{wd}} = 15\,000$. The factor decays linearly from one to zero over $n \in [0, N_{\text{wd}}/2]$ and remains zero thereafter, so that for $n \geq 7500$ the remaining Adam epochs and the entire L-BFGS stage are driven by the residual loss alone. This design provides an internal check against anchor-induced bias for exactly the channels where it matters most: the warm-down terminates with $\mathcal{L}_{\text{anchor}}=0$ identically, so the converged solution minimizes the driven-equation residual without any anchor contribution. When the warm-down schedule is in use, it is applied per channel automatically based on the source magnitude and requires no manual tuning. The threshold $s_a^*$ acts as a gate: channels with $\langle|V_{\text{short}} F_\ell|^2\rangle \geq s_a^*$ retain the static auto-anchor throughout training (no warm-down), while channels below the threshold engage the warm-down so that the converged solution is anchor-free. For the $^6$Li+$^{208}$Pb benchmark, the strong Coulomb suppression of $F_\ell$ pushes the absorption-to-transparency channels ($\ell \gtrsim 24$) below the threshold, where the warm-down is essential for an unbiased treatment of the near-transparent transition region. For the n+\nuc{40}{Ca} benchmark, low-$\ell$ channels ($\ell \leq 6$) sit above the threshold and use the static auto-anchor, while $\ell \geq 7$ fall just below it and engage the warm-down. The sensitivity scan presented next shows that the converged $|S_\ell|$ and $\delta_\ell$ are robust to this gating decision in either regime.

A more general convergence diagnostic is to vary the anchor parameters and verify that physical observables are insensitive to them. In practice this can be done by changing the anchor weight $w_a$ by a factor of a few or by varying $N_{\text{wd}}$, and confirming that $|S_\ell|$ and $\delta_\ell$ shift by amounts comparable to or below the residual disagreement of the benchmark with the reference solution, so that the anchor parameters are not the dominant source of error. For the present calculations this check is implicit in the warm-down construction itself, since for affected channels the late-stage solution is anchor-free regardless of $w_a$ and $s_a$. The cost of an explicit sensitivity scan is one additional training run per anchor parameter value, i.e.\ comparable to one extra partial wave per scan point. I have carried out this scan on two representative channels: varying $w_a$ across $\{0.05, 0.1, 0.2\}$ and $N_{\text{wd}}$ across $\{7500, 15\,000, 22\,500\}$ at $\ell = 25$ for the $^6$Li+$^{208}$Pb benchmark gives a spread of $3 \times 10^{-5}$ in $|S_{25}|$ and $8 \times 10^{-4}\,{}^\circ$ in $\delta_{25}$, two to four orders of magnitude below the residual $|\Delta|S||$ obtained against COLOSS at this $\ell$ (Sec.~\ref{sec:results}). An analogous $w_a$ scan on a representative weak-source channel of the n+\nuc{40}{Ca} benchmark ($j_{15/2}$, $\ell = 7$) gives $|S|$ and $\delta$ spreads of $4 \times 10^{-4}$ and $0.10^\circ$, comparable to the residual $\Delta\delta \approx 0.06^\circ$ against COLOSS at this $\ell$ (Sec.~\ref{sec:results}) and well below the worst-case $\Delta\delta = 0.60^\circ$ reported in Table~\ref{tab:n40Ca}.

The collocation points are distributed non-uniformly to concentrate sampling in regions of rapid variation. I allocate 50\% of the points to the nuclear interior $r \in [0, 2R_{\text{nuc}}]$, 20\% to the intermediate region $r \in [2R_{\text{nuc}}, R_0]$, and 30\% to the ECS region $r \in [R_0, R_{\text{max}}]$. Within each region, points are drawn from a uniform distribution. The required number of collocation points is set by the PINN optimization itself, not by a step-size criterion: each collocation point contributes one residual term to the loss, and an adequate density is needed everywhere to constrain the network output and prevent the gradient flow from finding spurious local minima of the loss. Empirically, reducing the n+\nuc{40}{Ca} budget below $\approx 1500$ points or the \nuc{6}{Li}+\nuc{208}{Pb} budget below $\approx 3000$ points degrades the high-$\ell$ phase shifts. The resulting average spacing (e.g.\ 0.015~fm for 4000 points in 60~fm) is therefore much smaller than the de~Broglie wavelength or the Woods-Saxon diffuseness ($a \approx 0.6$--0.8~fm); the latter scales would only require $\approx 100$ points to resolve, far fewer than the PINN optimization actually demands.

Training proceeds in two stages. The first stage uses the Adam optimizer~\cite{Kingma2015} with learning rate $5\times 10^{-4}$ and cosine annealing, where each epoch consists of one gradient descent step using all collocation points. Adam is robust to the choice of initial learning rate and handles the ill-conditioning of the loss landscape gracefully~\cite{Wang2021}. The second stage switches to L-BFGS~\cite{Liu1989} with strong Wolfe line search, where each epoch performs one quasi-Newton update with up to 20 line search evaluations. L-BFGS provides quadratic convergence near the minimum and refines the solution to high accuracy. Network weights are initialized with a fixed random seed (42) to ensure reproducibility. For systems with spin-orbit coupling, after training the $j = \ell - 1/2$ partner of a given $\ell$ I use its converged weights to initialize the $j = \ell + 1/2$ partner (transfer learning); since the two partners differ only in the spin-orbit term of the optical potential, this warm-start roughly halves the Adam and L-BFGS budgets needed by the second partner. For the spin-zero $^6$Li+$^{208}$Pb benchmark each $\ell$ has only a single channel and transfer learning is not used. All derivatives of the trial solution are computed via automatic differentiation of the network output with respect to the input coordinate, and training is accelerated on GPU hardware. The specific training parameters vary by system: for n+$^{40}$Ca, I use 2000 collocation points, 20\,000 Adam epochs, and 800 L-BFGS epochs; for $^6$Li+$^{208}$Pb, I use 4000 collocation points, 30\,000 Adam epochs, and 1000 L-BFGS epochs. For high angular momentum channels ($\ell \geq 6$, where the centrifugal barrier weakens the source term and slows convergence), the Adam and L-BFGS budgets for the initial training of each $\ell$ are increased by a factor of 1.5 (the spin-orbit partner trained by transfer learning uses the standard halved schedule and is not boosted). The increased computational requirements for the heavy-ion system arise from two factors: the large Sommerfeld parameter ($\eta \approx 15$) causes the Coulomb wave functions to vary more rapidly, requiring finer spatial sampling, and necessitates double-precision arithmetic throughout (rather than mixed precision), which increases the cost per epoch.

For heavy-ion systems with strong Coulomb barriers, I employ a causal training strategy~\cite{Wang2024} that builds the solution from the nuclear interior outward. The source term $V_N F_\ell$ in the driven equation is localized within the nuclear radius, so the scattered wave originates from this region and propagates outward. Training all radii simultaneously can lead to spurious solutions in the outer region where the source term vanishes. To avoid this, I progressively expand the training region from $R_{\text{nuc}}$ to $R_0$ over $N_{\text{warmup}}$ epochs (typically 3000):
\begin{equation}
L(n) = R_{\text{nuc}} + (R_0 - R_{\text{nuc}}) \cdot \min(n/N_{\text{warmup}}, 1),
\end{equation}
where $L(n)$ is the outer boundary of the training region at epoch $n$. Points beyond $L(n)$ contribute less to the loss, ensuring that the interior solution is established before fitting the asymptotic region.

\FloatBarrier
\section{Results}
\label{sec:results}

I validate the PINN-ECS method on two benchmark systems that test different aspects of the approach: nucleon-nucleus scattering with a complete optical potential including spin-orbit coupling, and heavy-ion scattering with strong Coulomb effects and a simpler (spin-independent) potential.

The first benchmark is neutron scattering on $^{40}$Ca at $E_{\text{lab}} = 20$~MeV, a well-studied system with extensive experimental data and established optical potentials. I use the Koning-Delaroche (KD02) global optical potential~\cite{Koning2003}, which writes the optical potential as the sum of four terms~\cite{Feshbach1958,Feshbach1962,Satchler1983}: a real volume term and an imaginary volume term, both proportional to a Woods-Saxon form factor~\cite{Woods1954} $f(x) = [1 + \exp((x-R)/a)]^{-1}$ with their own depths $V_v$, $W_v$, radii $R = r A_{\text{targ}}^{1/3}$, and diffuseness parameters; an imaginary surface term proportional to $-4 a_d\, \diff f/\diff x$ that peaks near the nuclear surface and accounts for the surface localization of low-energy absorption; and a spin-orbit term, also derivative of a Woods-Saxon form factor, that couples the projectile spin to its orbital angular momentum and splits each $\ell > 0$ partial wave into $j = \ell + 1/2$ and $j = \ell - 1/2$ components. The Koning-Delaroche parametrization provides energy- and target-dependent values for all these components; similar global parametrizations include the Chapel Hill potential~\cite{Varner1991}. I compute 21 channels in total, corresponding to $\ell = 0$ through $\ell = 10$: $s_{1/2}$ for $\ell = 0$, and both $j = \ell \pm 1/2$ spin-orbit partners for $\ell = 1$--$10$.

For the ECS transformation I use $R_0 = 20$~fm, $R_{\text{max}} = 60$~fm, $\theta = 12^\circ$, and a transition width $w = 0.1 R_0 = 2$~fm. These choices follow the standard criteria established in the atomic and molecular ECS literature~\cite{McCurdy2004,Rescigno2000}: $R_0$ must exceed the range of the nuclear potential (the Woods-Saxon half-value radius is $R_V \approx 4.3$~fm here, so $R_0 = 20$~fm is comfortably beyond it); $\theta$ must be large enough that outgoing waves are adequately damped, but small enough that the Coulomb potential $1/x(r)$ at complex coordinates does not amplify round-off errors and the rapidly decaying solution can still be resolved by a moderate number of collocation points; and $R_{\text{max}}$ must give a small ECS damping factor $\exp[-k(R_{\text{max}}-R_0)\sin\theta]$, which evaluates to $\approx 4 \times 10^{-4}$ for the present parameters and is well below the $10^{-3}$ threshold required for adequate absorption at the outer boundary. The same general criteria, applied to the heavier system, motivate the slightly different parameters quoted below for $^6$Li+$^{208}$Pb.

\begin{figure}[t]
\centering
\includegraphics[width=\columnwidth]{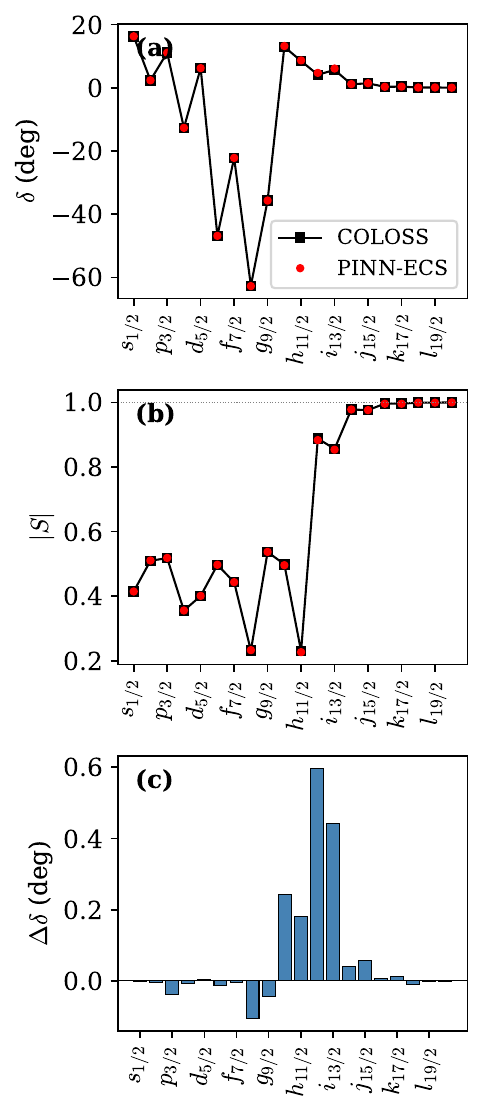}
\caption{PINN-ECS results for n+$^{40}$Ca scattering at $E_{\text{lab}} = 20$~MeV with the KD02 optical potential. (a) Phase shifts $\delta$ comparing PINN-ECS (circles) to COLOSS reference (squares). (b) S-matrix magnitudes $|S|$. (c) Phase shift errors $\Delta\delta = \delta_{\text{PINN}} - \delta_{\text{COLOSS}}$.}
\label{fig:n40Ca_phases}
\end{figure}

\begin{figure}[t]
\centering
\includegraphics[width=\columnwidth]{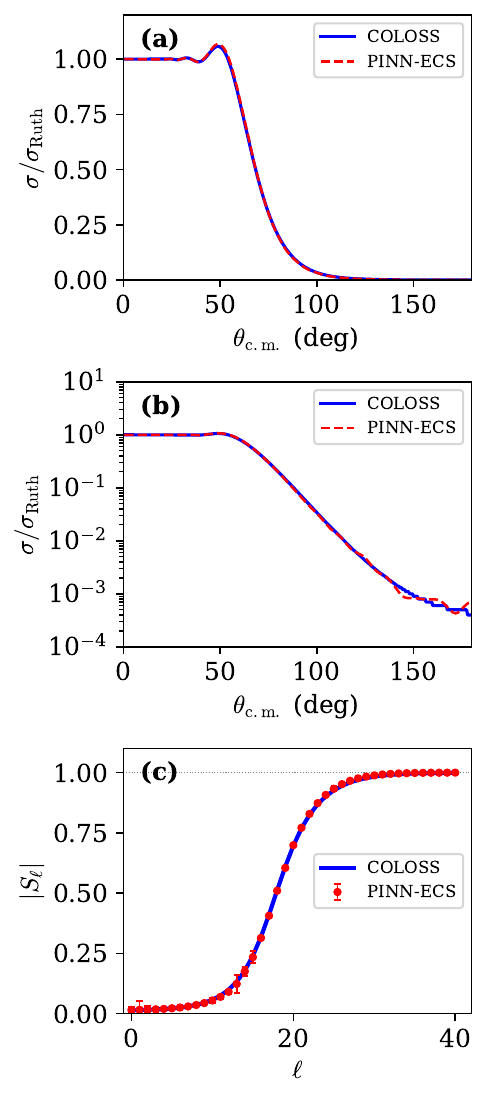}
\caption{$^6$Li+$^{208}$Pb elastic scattering at $E_{\text{lab}} = 40$~MeV. (a) Rutherford ratio on a linear scale. (b) Rutherford ratio on a logarithmic scale, showing the diffractive structure over three orders of magnitude. (c) S-matrix magnitude $|S_\ell|$ versus partial wave $\ell$; error bars show the standard deviation across five matching radii ($r_{\rm match} = 14$--$18$~fm). In all panels, COLOSS is the solid line and PINN-ECS is the dashed line or circles.}
\label{fig:6Li208Pb}
\end{figure}

Figure~\ref{fig:n40Ca_phases} compares PINN-ECS results with reference values from the COLOSS code~\cite{Liu2025}, which solves the same scattering problem using conventional complex scaling with Lagrange-mesh discretization. Panel~(a) shows the phase shifts $\delta$ for all 21 channels, where black squares connected by lines represent the COLOSS reference and red circles show the PINN-ECS results. The two methods agree so well that the symbols overlap for most channels. Panel~(b) displays the S-matrix magnitudes $|S|$, which characterize absorption into non-elastic channels. The horizontal dashed line marks the unitarity limit $|S| = 1$. Panel~(c) shows the phase shift errors $\Delta\delta = \delta_{\text{PINN}} - \delta_{\text{COLOSS}}$ as a bar chart. For the strongly absorbed channels ($\ell \leq 4$), the errors are at the $0.1^\circ$ level. For the higher partial waves ($\ell = 5$--$10$), the errors increase but remain $\leq 0.60^\circ$ in all cases, with the worst case at $i_{11/2}$ ($\ell = 6$). The mean $|\Delta\delta|$ across all 20 spin-orbit channels is $0.09^\circ$. The matching-radius spread (the analogue of the error bars shown in Fig.~\ref{fig:6Li208Pb}(c) for the heavy-ion benchmark) is below $0.08^\circ$ for all 21 channels (mean $0.012^\circ$) and would be smaller than the symbol size in Fig.~\ref{fig:n40Ca_phases}; it is therefore omitted from the plot.

Table~\ref{tab:n40Ca} presents a quantitative comparison for representative channels spanning the full angular momentum range. The $s_{1/2}$ wave shows $\delta_{\text{PINN}} = 16.27^\circ$ compared to $\delta_{\text{COLOSS}} = 16.27^\circ$, agreement to two decimal places. The $f_{5/2}$ wave, with a large negative phase shift of $-46.86^\circ$, is reproduced to within $0.01^\circ$. The most challenging channel is $i_{11/2}$ ($\ell = 6$, $j = 11/2$), where the PINN phase shift of $4.70^\circ$ overestimates the COLOSS value of $4.10^\circ$ by $\Delta\delta = 0.60^\circ$; the $j_{15/2}$ channel ($\ell = 7$) for which the auto-adaptive logic of Sec.~\ref{sec:method} engages the warm-down agrees with COLOSS to $\Delta\delta = 0.06^\circ$. High-$\ell$ channels have small source terms due to centrifugal suppression, making accurate training more difficult, but the S-matrix magnitudes still agree with COLOSS to within $\Delta|S| \leq 0.005$ across all 20 spin-orbit channels, confirming that the imaginary part of the optical potential is correctly handled.

\begin{table}[t]
\caption{Phase shifts and S-matrix elements for n+$^{40}$Ca at $E_{\text{lab}} = 20$~MeV. Representative channels are shown from the interior ($s_{1/2}$--$f_{5/2}$), surface ($g_{9/2}$--$h_{11/2}$), and high angular momentum ($i_{11/2}$--$j_{15/2}$) regions, including the channel with the worst agreement ($i_{11/2}$, $\Delta\delta = 0.60^\circ$).}
\label{tab:n40Ca}
\begin{ruledtabular}
\begin{tabular}{lrrrrr}
Channel & $\delta_{\text{PINN}}$ & $\delta_{\text{COL}}$ & $\Delta\delta$ & $|S|_{\text{PINN}}$ & $|S|_{\text{COL}}$ \\
\hline
$s_{1/2}$ & $+16.27^\circ$ & $+16.27^\circ$ & $+0.00^\circ$ & 0.415 & 0.415 \\
$p_{3/2}$ & $+11.06^\circ$ & $+11.10^\circ$ & $-0.04^\circ$ & 0.519 & 0.518 \\
$d_{5/2}$ & $+6.19^\circ$ & $+6.19^\circ$ & $+0.00^\circ$ & 0.401 & 0.401 \\
$f_{5/2}$ & $-46.86^\circ$ & $-46.85^\circ$ & $-0.01^\circ$ & 0.498 & 0.497 \\
$g_{9/2}$ & $-35.70^\circ$ & $-35.66^\circ$ & $-0.04^\circ$ & 0.537 & 0.537 \\
$h_{11/2}$ & $+8.64^\circ$ & $+8.46^\circ$ & $+0.18^\circ$ & 0.227 & 0.230 \\
$i_{11/2}$ & $+4.70^\circ$ & $+4.10^\circ$ & $+0.60^\circ$ & 0.883 & 0.888 \\
$j_{15/2}$ & $+1.44^\circ$ & $+1.38^\circ$ & $+0.06^\circ$ & 0.976 & 0.976 \\
\end{tabular}
\end{ruledtabular}
\end{table}

\begin{figure}[t]
\centering
\includegraphics[width=\columnwidth]{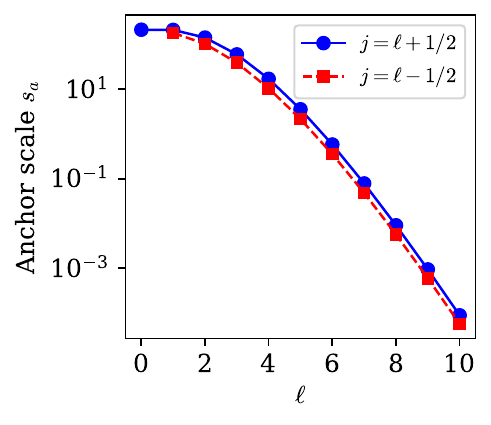}
\caption{Auto-adaptive anchor scale $s_a = \langle|V_{\text{short}} F_\ell|^2\rangle$ versus angular momentum $\ell$ for n+$^{40}$Ca. The scale decreases by over three orders of magnitude from $\ell = 0$ to $\ell \approx 6$ due to centrifugal suppression of the source term, and reaches the threshold $s_a^* = 0.1$ near $\ell = 7$. Circles with solid line: $j = \ell + 1/2$; squares with dashed line: $j = \ell - 1/2$. The auto-adaptive logic of Sec.~\ref{sec:method} retains the static auto-anchor for $\ell \leq 6$ (above threshold) and engages the warm-down for $\ell \geq 7$ (below threshold); the sensitivity scan reported in Sec.~\ref{sec:method} confirms that the converged $|S_\ell|$ and $\delta_\ell$ are robust to this gating decision.}
\label{fig:auto_anchor}
\end{figure}

The auto-adaptive anchor mechanism plays a role in achieving consistent accuracy across partial waves. Figure~\ref{fig:auto_anchor} shows the anchor scale $s_a$ for each partial wave of the n+\nuc{40}{Ca} system: it drops by over three orders of magnitude from $\ell = 0$ to $\ell \approx 6$ due to centrifugal suppression of the source term $V_N F_\ell$, and reaches the threshold $s_a^* = 0.1$ near $\ell = 7$. The auto-adaptive logic of Sec.~\ref{sec:method} therefore retains the static auto-anchor for $\ell \leq 6$ and engages the warm-down for $\ell \geq 7$; the sensitivity scan reported in Sec.~\ref{sec:method} confirms that the converged S-matrix is robust to this gating decision.

Beyond the phase shifts and S-matrix elements, it is instructive to compare the wave functions themselves, as these are relevant for applications to nuclear reactions beyond elastic scattering.  Table~\ref{tab:wfn_mse} reports the relative RMS difference between PINN and Runge-Kutta reference-solver wave functions for representative n+\nuc{40}{Ca} channels, evaluated in three radial regions: the nuclear interior ($r < R_{\rm nuc}$), the nuclear surface ($R_{\rm nuc} < r < 2R_{\rm nuc}$), and the outer region ($r > 2R_{\rm nuc}$). For low-$\ell$ channels ($\ell \leq 3$), the relative RMS is below $0.15\%$ in all regions, and below $0.05\%$ for all spin-orbit partners except $p_{3/2}$. The worst case across the higher partial waves is $i_{11/2}$ ($\ell = 6$, the same channel as the worst-case $\Delta\delta$ of Table~\ref{tab:n40Ca}), at $1.4\%$ in the interior, $1.2\%$ at the surface, and $1.6\%$ in the outer region. The auto-adaptive warm-down for $\ell \geq 7$ leaves the wave-function quality at the $0.1\%$ level even at the high-angular-momentum tail (e.g.\ $j_{15/2}$: $0.04\%$ interior, $0.03\%$ surface, $0.11\%$ outer). The mean relative RMS across all 21 channels is $0.20\%$, confirming that the PINN solution reproduces the wave function accurately in the region most relevant for nuclear reaction calculations.

\begin{table}[t]
\caption{Relative RMS difference between PINN and Runge-Kutta reference-solver wave functions for n+$^{40}$Ca, evaluated in the nuclear interior ($r < R_{\rm nuc}$), surface ($R_{\rm nuc} < r < 2R_{\rm nuc}$), and outer ($r > 2R_{\rm nuc}$) regions.}
\label{tab:wfn_mse}
\begin{ruledtabular}
\begin{tabular}{lrrrr}
Channel & Interior & Surface & Outer & Total \\
\hline
$s_{1/2}$  & 0.03\% & 0.02\% & 0.02\% & 0.02\% \\
$f_{5/2}$  & 0.02\% & 0.02\% & 0.04\% & 0.03\% \\
$g_{9/2}$  & 0.06\% & 0.05\% & 0.07\% & 0.06\% \\
$h_{11/2}$ & 0.14\% & 0.25\% & 0.42\% & 0.33\% \\
$i_{11/2}$ & 1.42\% & 1.16\% & 1.56\% & 1.39\% \\
$j_{15/2}$ & 0.04\% & 0.03\% & 0.11\% & 0.08\% \\
\end{tabular}
\end{ruledtabular}
\end{table}

The second benchmark is $^6$Li+$^{208}$Pb elastic scattering at $E_{\text{lab}} = 40$~MeV, a heavy-ion system with strong Coulomb effects. The Sommerfeld parameter for this system is $\eta \approx 15$, indicating that multiple Coulomb scattering events occur before the nuclear interaction, creating a thick Coulomb barrier that strongly modifies the scattering pattern. I compute 41 partial waves from $\ell = 0$ to $\ell = 40$ to capture the complete angular distribution. In this calculation, \nuc{6}{Li} is treated as a spin-0 projectile; spin-orbit coupling is not included. The optical potential is a Woods-Saxon volume form with parameters $V_v = 109.5$~MeV, $r_v = 1.326$~fm, $a_v = 0.811$~fm, $W_v = 22.384$~MeV, $r_w = 1.534$~fm, $a_w = 0.884$~fm, taken from a fit to elastic scattering data in this mass and energy region~\cite{Cook1982}. The Coulomb potential is that of a uniformly charged sphere with radius $R_C = r_C A_{\rm targ}^{1/3}$ using $r_C = 1.3$~fm, giving $R_C \approx 7.7$~fm. The ECS parameters are $R_0 = 20$~fm, $R_{\text{max}} = 50$~fm, and $\theta = 12^\circ$; the much larger wave number ($k \approx 3.29$~fm$^{-1}$) produces a damping factor of approximately $10^{-9}$ at the outer boundary, far smaller than the $10^{-3}$ threshold of Sec.~\ref{sec:theory} and more than sufficient even with a smaller $R_{\text{max}}$ than for n+\nuc{40}{Ca}. For this heavy-ion system, I employ the causal training strategy with $N_{\text{warmup}} = 3000$ epochs to ensure stable convergence across all partial waves.

Figure~\ref{fig:6Li208Pb} presents the results for the $^6$Li+$^{208}$Pb system. Panel~(a) shows the Rutherford ratio on a linear scale, which exhibits oscillations arising from interference between the Coulomb and nuclear scattering amplitudes, a signature feature of heavy-ion scattering near the Coulomb barrier. Panel~(b) displays the same quantity on a logarithmic scale, revealing the diffractive structure over three orders of magnitude; at backward angles ($\theta > 140^\circ$) the PINN-ECS curve shows slight fluctuations at the $10^{-3}$ level, reflecting the accumulated effect of small S-matrix errors at high $\ell$. Panel~(c) shows the S-matrix magnitudes with error bars indicating the standard deviation across five matching radii. The magnitudes exhibit the characteristic pattern of heavy-ion scattering: strong absorption ($|S| \ll 1$) at low angular momenta where the trajectories penetrate inside the Coulomb barrier, a grazing region around $\ell \approx 15$--25 where the absorption falls off rapidly, and transparency ($|S| \to 1$) at high angular momenta where the centrifugal barrier prevents nuclear contact. The PINN-ECS results agree well with the COLOSS reference for all three observables.

For the n+\nuc{40}{Ca} system, I present individual phase shifts because they provide a channel-by-channel diagnostic of the method's accuracy for each partial wave. For \nuc{6}{Li}+\nuc{208}{Pb}, I instead show the Rutherford ratio and S-matrix magnitudes because the elastic cross section is the standard experimental observable for heavy-ion scattering and provides a stringent test of the coherent sum over 41 partial waves. The underlying phase shifts for \nuc{6}{Li}+\nuc{208}{Pb} are fully consistent with the S-matrix comparison: PINN-ECS and COLOSS phase shifts agree to within $0.13^\circ$ throughout the transition region ($\ell = 25$--$30$) and to better than $0.6^\circ$ for $\ell = 13$--$24$. For the strongly absorbed channels ($\ell \leq 12$, $|S_\ell| \lesssim 0.1$), the phase shift becomes ill-conditioned because $\delta_\ell = \arg(S_\ell)/2$ is highly sensitive to small variations in $S_\ell$ when $|S_\ell|$ is small; the PINN-COLOSS phase shift difference can reach $\approx 1.8^\circ$ in this regime, but the corresponding $|\Delta|S_\ell||$ remains below $0.002$ and the contribution to the elastic cross section is negligible.

Table~\ref{tab:6Li208Pb} presents a quantitative comparison of S-matrix elements for selected partial waves spanning the absorption-to-transparency transition. The agreement between PINN-ECS and COLOSS is excellent across the entire range, with $|S|$ differences below 0.01 in the absorption region ($\ell \lesssim 20$) and $\Delta|S| \lesssim 0.007$ even in the most challenging transition region ($\ell = 25$--30). The mean $|\Delta|S||$ across all 41 partial waves is $3.3 \times 10^{-3}$. The accurate treatment of the transition region is enabled by the auto-adaptive anchor warm-down described in Sec.~\ref{sec:method}: as the source term magnitude $\langle|V_{\text{short}} F_\ell|^2\rangle$ falls below $\approx 10^{-1}$ for $\ell \gtrsim 24$, the warm-down progressively disables the anchor loss in the second half of Adam training, removing the bias that a fixed anchor scale would otherwise impose on near-transparent channels where the correct scattered wave is itself very small.

To quantify the numerical uncertainty, I extracted the S-matrix at each of the five matching radii individually and computed the standard deviation, shown as error bars in Fig.~\ref{fig:6Li208Pb}(c). The spread remains below $\approx 10^{-2}$ in the transition region ($\ell = 25$--$30$) and below $0.001$ for well-converged channels ($\ell \geq 30$); it is largest (up to $\approx 4 \times 10^{-2}$) near the absorption shoulder ($\ell \approx 13$), where the small $|S_\ell|$ amplifies the apparent matching-radius variation while the absolute $S$-matrix differences against COLOSS stay below $0.005$.

\begin{table}[t]
\caption{S-matrix elements for $^6$Li+$^{208}$Pb at $E_{\text{lab}} = 40$~MeV. Selected partial waves are shown comparing PINN-ECS to the COLOSS reference.}
\label{tab:6Li208Pb}
\begin{ruledtabular}
\begin{tabular}{lrrrrr}
$\ell$ & $\text{Re}\,S_{\text{PINN}}$ & $\text{Re}\,S_{\text{COL}}$ & $\text{Im}\,S_{\text{PINN}}$ & $\text{Im}\,S_{\text{COL}}$ & $\Delta|S|$ \\
\hline
0  & $-0.014$ & $-0.014$ & $-0.007$ & $-0.006$ & $+0.001$ \\
5  & $-0.012$ & $-0.011$ & $-0.019$ & $-0.019$ & $+0.000$ \\
10 & $+0.035$ & $+0.039$ & $-0.042$ & $-0.042$ & $-0.003$ \\
15 & $+0.231$ & $+0.237$ & $+0.037$ & $+0.037$ & $-0.005$ \\
20 & $+0.687$ & $+0.684$ & $+0.129$ & $+0.130$ & $+0.002$ \\
25 & $+0.932$ & $+0.925$ & $+0.040$ & $+0.044$ & $+0.007$ \\
30 & $+0.989$ & $+0.983$ & $+0.007$ & $+0.009$ & $+0.005$ \\
40 & $+1.000$ & $+0.999$ & $+0.000$ & $+0.000$ & $+0.001$ \\
\end{tabular}
\end{ruledtabular}
\end{table}

The Rutherford ratio shown in Fig.~\ref{fig:6Li208Pb}(a,b) is computed from the S-matrix elements via the standard partial-wave expansion of the elastic amplitude into a Coulomb part and a nuclear-induced part~\cite{Frobrich1996,Thompson2009}. With 41 partial waves the angular distribution is well-converged at all angles shown.

\begin{figure}[t]
\centering
\includegraphics[width=\columnwidth]{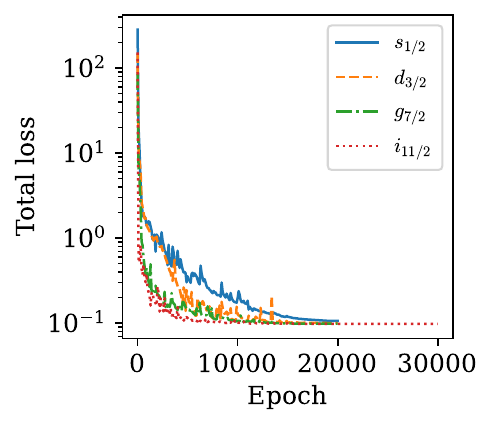}
\caption{Training convergence for n+$^{40}$Ca scattering. Four representative channels are shown: $s_{1/2}$ (solid), $d_{3/2}$ (dashed), $g_{7/2}$ (dash-dotted), and $i_{11/2}$ (dotted). The initial loss for low-$\ell$ channels is significantly larger than for high-$\ell$ channels, primarily because of the centrifugal suppression of the source term $V_N F_\ell$. All channels then drop by two to three orders of magnitude during the Adam stage and approach a similar floor near $\approx 10^{-1}$ that reflects the irreducible residual after the anchor contribution has decayed.  Despite this finite floor, all four channels yield phase shifts accurate to within $0.60^\circ$ (see Fig.~\ref{fig:n40Ca_phases}), showing that the optimizer has converged to physically correct solutions even where the loss does not drop further.}
\label{fig:convergence}
\end{figure}

Figure~\ref{fig:convergence} illustrates the training dynamics for the n+$^{40}$Ca benchmark. Channels are trained sequentially, with transfer learning applied within each $\ell$ value (from $j=\ell-1/2$ to $j=\ell+1/2$). The loss decreases rapidly during training, dropping by several orders of magnitude. Notably, the $s_{1/2}$ channel converges to a relatively high loss ($\approx 0.8$) compared to other channels ($\approx 0.1$), yet still produces accurate phase shifts and S-matrix elements (see Fig.~\ref{fig:n40Ca_phases}). This demonstrates the robustness of the PINN-ECS method: the physical observables are insensitive to the precise value of the training loss once the solution has captured the essential features of the wave function. Each partial wave requires on the order of a few minutes on a modern GPU.

\FloatBarrier
\section{Discussion}
\label{sec:discussion}

The results presented above demonstrate that the combination of exterior complex scaling with physics-informed neural networks provides an accurate and reliable method for solving nuclear scattering problems. The key enabling insight is that ECS transforms the oscillatory scattering boundary conditions into decaying waves that neural networks can represent naturally, while the driven-equation formulation confines the nuclear potential to the real axis where it is well-defined. Several aspects of the method merit further discussion.

The accuracy achieved for nucleon-nucleus scattering is comparable to that of established numerical methods. For the strongly absorbed channels ($\ell \leq 4$), phase shift errors are at the $0.1^\circ$ level. The errors increase modestly for higher partial waves, reaching $\Delta\delta = 0.60^\circ$ for the worst channel ($i_{11/2}$), due to the combination of weak source terms (centrifugal suppression) and the sensitivity of small phase shifts to numerical noise. The mean $|\Delta\delta|$ across all 20 spin-orbit channels is $0.09^\circ$. A detailed wave function comparison (Table~\ref{tab:wfn_mse}) shows that the relative RMS difference between PINN and Runge-Kutta reference-solver solutions is below $0.15\%$ for $\ell \leq 3$ and below $2\%$ even for the worst case, with the interior accuracy typically better than the outer region (the high-$\ell$ near-transparent channels are the exception, where the absolute scale of the wave function in the interior is centrifugally suppressed and the relative metric is dominated by noise). For the heavy-ion benchmark, the auto-adaptive anchor warm-down maintains $\Delta|S| \lesssim 0.007$ even in the absorption-to-transparency transition region ($\ell = 25$--30) where the scattered wave amplitude is smallest, and the mean $|\Delta|S||$ across all 41 partial waves is $3.3 \times 10^{-3}$, with no additional training cost from the warm-down itself beyond the standard $1.5\times$ epoch boost already applied to all $\ell \geq 6$ channels. The interior wavefunction accuracy reported here suggests that the PINN-ECS solutions should be of useful quality for inelastic and transfer applications where the wave function inside the nuclear volume enters the transition amplitude, although a quantitative test in those reaction channels is left to future work; if higher accuracy is needed it can be sought through finer collocation near the nuclear surface or by extended training of the highest-$\ell$ channels (with the algorithmic choice between static-floor anchor and warm-down made according to whether the system is dominated by short-range absorption or by a strong-Coulomb absorption-to-transparency transition, respectively).

The driven-equation approach has a significant advantage over directly solving for the total wave function: it naturally separates the known asymptotic behavior (encoded in the incident Coulomb wave) from the unknown scattered contribution. Although the scattered wave oscillates with the same wave number $k$ as the incident wave in the nuclear interior, it is exponentially damped in the ECS region ($r > R_0$) by construction, while the total wave function would continue oscillating. The neural network therefore only needs to represent a function that decays to zero at large $r$, which simplifies the learning task. Furthermore, the source term $V_{\text{short}} F_\ell$ provides an automatic scale for the solution magnitude, reducing the ambiguity inherent in homogeneous eigenvalue problems.

The applicability of the PINN-ECS method is limited to reactions where the relevant physics is confined to radii smaller than the ECS scaling radius $R_0$. This includes elastic scattering, inelastic excitation, and transfer reactions driven by short-range nuclear interactions. However, processes dominated by long-range electromagnetic transitions, such as Coulomb excitation at sub-barrier energies where the excitation probability depends on the wave function at large impact parameters ($r \gg R_0$), cannot be treated directly within this framework because the ECS transformation modifies the wave function precisely in the region where the electromagnetic coupling is strongest. For such processes, alternative approaches such as semiclassical Coulomb excitation theory~\cite{Alder1975} or explicit treatment of the long-range coupling before applying ECS would be needed.

The computational cost of the PINN-ECS method is dominated by the training iterations required for each partial wave. On a modern GPU, each channel requires on the order of a few minutes, so a complete calculation with 20--40 partial waves takes roughly one to two hours of wall-clock time. This is significantly longer than conventional integration methods (which require seconds per partial wave) but acceptable for research applications where end-to-end differentiability or mesh-free formulations are required. The training time could be further reduced by using smaller networks for simple potentials or by warm-starting from nearby energies~\cite{Goswami2020,Chakraborty2021}.

\FloatBarrier
\section{Conclusions}
\label{sec:conclusions}

I have demonstrated that exterior complex scaling enables physics-informed neural networks to solve nuclear reaction problems. The key innovation is using ECS to transform oscillatory scattering boundary conditions into exponentially decaying waves that neural networks can represent naturally, combined with a driven-equation formulation that confines nuclear potentials to the real axis. The method has been validated on nucleon-nucleus scattering (n+$^{40}$Ca at 20~MeV) achieving phase shift accuracy of $\Delta\delta \lesssim 0.1^\circ$ for channels with $\ell \leq 4$ and $\Delta\delta \leq 0.60^\circ$ for all 20 spin-orbit channels (mean $|\Delta\delta| = 0.09^\circ$), with wave function accuracies below $0.15\%$ for low-$\ell$ channels, and on heavy-ion scattering ($^6$Li+$^{208}$Pb at 40~MeV) where the mean $|\Delta|S||$ across all 41 partial waves is $3.3 \times 10^{-3}$ and the Rutherford ratio reproduces the COLOSS reference over three orders of magnitude, with deviations only at the $10^{-3}$ level at backward angles where the cross section is smallest.

Several technical innovations contribute to the reliability of the method: the sigmoid-capped boundary condition factor prevents high-$\ell$ pathologies by limiting the growth of the angular momentum prefactor; the auto-adaptive anchor scale keeps the residual and anchor losses comparable across all partial waves, with a per-channel schedule selected automatically based on the source magnitude (the static auto-anchor for strongly-sourced channels, switching to a linear warm-down for weak-source channels where the anchor would otherwise bias $|S_\ell|$ below unity, including the absorption-to-transparency transition of $^6$Li+$^{208}$Pb and the high-$\ell$ tail of n+\nuc{40}{Ca}); and the two-stage training protocol (Adam followed by L-BFGS) provides both robust initial convergence and high final accuracy. The extraction of phase shifts using automatic differentiation and multi-point S-matrix averaging ensures that derivatives are computed accurately without numerical finite-difference errors.

The PINN-ECS approach opens several directions for future research, building on the broader trend of integrating machine learning into nuclear physics~\cite{Boehnlein2022}. The most immediate application is inverse problems: fitting optical potential parameters directly to experimental elastic scattering data by exploiting the end-to-end differentiability of the computational pipeline. Coupled-channels calculations for reactions involving particle transfer or nuclear excitation~\cite{Thompson1988,Austern1987,Hagino2022} could be addressed by representing multiple channel wave functions with a multi-output network, avoiding the need to explicitly construct coupled integro-differential equations. Extension to few-body scattering problems~\cite{Deltuva2005,Lazauskas2018b,Lazauskas2019Symphony,Lazauskas2019}, where the mesh-free nature of neural networks may help mitigate the curse of dimensionality faced by grid-based methods, represents a longer-term goal.

More broadly, this work demonstrates that the combination of physics-informed machine learning with analytic transformations such as complex scaling can extend the reach of neural network methods to problems that appear intractable within the standard PINN framework. The philosophy of transforming a difficult problem into one that matches the inductive biases of neural networks, namely smooth, bounded, decaying solutions, may prove fruitful in other areas of computational physics where boundary conditions pose fundamental challenges.

\begin{acknowledgments}
This work was supported by the National Natural Science Foundation of China (Grant Nos.~12475132 and 12535009) and the Fundamental Research Funds for the Central Universities. In preparing this work I used large language models to assist with code development and language editing; all physical results, conclusions, and the final text are my own and were checked by me.
\end{acknowledgments}

\FloatBarrier
\bibliography{references}

\end{document}